\documentclass[aps,prc,showpacs,twocolumn,preprintnumbers,floatfix,
showkeys,tightenlines]{revtex4}
\usepackage{amsmath,amssymb,amsfonts}
\usepackage{graphicx}
\usepackage{color}
\allowdisplaybreaks

\newcommand{\be}{\begin{equation}}
\newcommand{\ee}{\end{equation}}
\newcommand{\bea}{\begin{eqnarray}}
\newcommand{\eea}{\end{eqnarray}}
\newcommand{\bm}{\bibitem}

\newcommand{\ep}{\epsilon}

\newcommand{\om}{\omega}

\newcommand{\lm}{\lambda}

\newcommand{\sg}{\sigma}

\newcommand{\ze}{\zeta}

\newcommand{\cz}{{\cal Z}}

\begin{document}



\title{Symmetry coefficients and incompressibility of clusterized
supernova matter}

\author{S.K. \surname{Samaddar}${^1}$}
\email{santosh.samaddar@saha.ac.in}
\author{J.N. \surname{De}${^1}$}
\email{jn.de@saha.ac.in}
\author{X. \surname{Vi\~nas}$^{2}$}
\email{xavier@ecm.ub.es}
\author{M. \surname{Centelles}$^{2}$}
\email{mariocentelles@ub.edu}
\affiliation{
$^1$Saha Institute of Nuclear Physics, 1/AF Bidhannagar, Kolkata
{\sl 700064}, India \\
$^2$Departament d'Estructura i Constituents de la Mat\`eria,
Facultat de F\'{\i}sica, \\
and Institut de Ci\`encies del Cosmos, Universitat de Barcelona, \\
Diagonal {\sl 647}, {\sl 08028} Barcelona, Spain}


\begin{abstract} 
The symmetry energy coefficients, incompressibility, 
and  single-particle and isovector potentials  of clusterized dilute
nuclear matter are calculated at different temperatures employing the
$S$-matrix approach to the evaluation of the equation of state.
Calculations have been extended to understand the aforesaid properties
of homogeneous and clusterized supernova matter in the subnuclear
density region. Comparison of the results 
in the $S$-matrix and mean-field approach reveals
some subtle differences in the density and temperature region we
explore.
\end{abstract}


\maketitle

\section{Introduction}

Studies of the dynamical evolution of supernovae require an accurate
knowledge of the properties of nuclear matter over a large range of
densities $\rho $, temperatures $T$, and proton concentrations $Y_p$
\cite{bet,jan}. The density values range from a few times the normal
nuclear matter density $\rho_0 $ to $\sim 10^{-7} \rho_0 $, the
temperature may be as high as $\sim$ 20 MeV, and the proton
concentration may even be close to zero. Based on the laboratory
experiments, the properties of cold ($T=0$) nuclear matter around the
saturation density $\rho_0 $ with proton concentration limited by the
stability valley of finite nuclei are more or less well known.
Outside this narrow limit, the understanding of the
properties of nuclear matter is fraught with many uncertainties.

The density dependence of the nuclear incompressibility and of the
symmetry energy are among the key elements in the simulation dynamics
of supernova explosion \cite{ste}. At supranormal densities, nuclear
flow in energetic nuclear collisions aided by model-dependent
calculations helps to understand the nuclear incompressibility
\cite{sto,dan02,bar05}. Similarly, inference can be made on the
density dependence of the symmetry energy from the comparison of
theoretical predictions with experimental data on the differential
flow of neutrons and protons and from the $\pi^-/\pi^+$, $K^0/K^+$
ratios, etc \cite{li,fer,li09}. Disassembly of hot expanded nuclei
offers one of the best tools to study the characteristics of the
nuclear symmetry energy at subnormal densities \cite{she,kow,sam}.
Experimental data related to isotopic distributions \cite{bot},
isospin diffusion \cite{tsa,che,tsa09}, and isoscaling \cite{she,sou}
constrain the density dependence of the symmetry energy in the
subnormal region. There is considerable uncertainty, however, in all
these extractions. The importance of the nuclear symmetry energy can
be gauged by the fact that higher symmetry energy, for example, leads
to a lower electron ($e^-$)-capture rate in the supernova collapse
phase that may result in a strong explosive shock \cite{bet,sum}. The
isotopic abundance of relatively heavier elements in explosive
nucleosynthesis is further directly correlated to the symmetry energy.

Microscopic approaches based on realistic $NN$ interactions, Brueckner
or variational schemes, or on effective field theories show a large
range of predictions on the density dependence of the nuclear
incompressibility \cite{ban,cha,piek04,colo04,dex08,piek09} and symmetry
energy \cite{li,hei00,fur02,fuc,xu,cen09}. These results are model
dependent, which seems unavoidable at near-normal densities and above.
At lower densities, however, based on 
the general analysis of the grand-canonical
partition function for nuclear matter in the $S$-matrix framework
\cite{mal,de}, it is possible to have predictions for various nuclear
observables which are practically model independent.
For dilute nuclear matter, the system minimizes its total free energy
by forming clusters. The observables are expressed in terms of
specific known properties of these cluster species like their binding
energies, scattering phase shifts, etc. Using the virial expansion
technique, Horowitz and Schwenk \cite{hor} have evaluated the symmetry
energy coefficients of clusterized dilute nuclear matter where the
cluster species were neutrons, protons, and $\alpha $ particles. In
the $S$-matrix approach, the calculations in nuclear matter were
extended with inclusion of all possible heavier clusters \cite{mal}.
The so-calculated results were found to be appreciably different from
those obtained in Ref.\ \cite{hor}. 

In charge-free nuclear matter, the fragment species, in principle, may
be infinite in size.
In supernova matter (charge-neutral due to the presence of electrons),
the Coulombic term in the binding energies of the fragments severely
constrains their size within the limits of the drip lines. The
fragment composition is then likely to be altered, which would affect
the properties of supernova matter in contrast to those in nuclear
matter. The aim of the present paper is to investigate the symmetry
and compression properties of supernova matter in the low-density
regime (up to $\sim $ 0.02 fm$^{-3}$) along with those for nuclear
matter. Specifically, the density and temperature dependence of the
symmetry energy coefficients, the isovector potentials, and the
symmetry incompressibilities are explored.

The paper is organised as follows. The theoretical details are presented
in Sec.~II. The results and discussions are contained in Sec.~III, and
the concluding remarks are given in Sec.~IV.

\section{Elements of theory}

The logical framework of the theory is set in the grand-canonical
partition function of the interacting quantum system in the $S$-matrix
formalism of statistical mechanics as proposed by Dashen, Ma, and
Bernstein \cite{das}. In subsection II.A, we present the key elements
of the theory \cite{mal} as applied to nuclear matter or supernova
matter. In subsection II.B, the expressions for a few relevant
observables are given. In subsection II.C we present the methodology
for calculating the  
single-particle and isovector potentials
 and the symmetry coefficients in some detail.

\subsection{The $S$-matrix framework}

The grand-canonical partition function for the two-component nuclear
matter composed of neutrons and protons as the elementary species
is written as
\be
\cz = {\rm Tr}\, e^{-\beta (H-\mu_p\hat{N}_p -\mu_n\hat{N}_n)}\,,
\ee
where $\beta $ is the inverse of the temperature $T$
of the system, $H$ the total
Hamiltonian, $\hat{N}_{p,n}$ the number operators for protons and neutrons,
and $\mu_{p,n}$ are the corresponding chemical potentials. The trace
is taken over any complete set of states of all possible number of 
nucleons. Denoting the elementary 
fugacities by $\zeta_p=e^{\beta \mu_p}$ and
$\zeta_n=e^{\beta \mu_n}$, the full trace can be decomposed as
\be
\cz = \sum_{Z,N=0}^\infty (\ze_p)^Z (\ze_n)^N 
\, {\rm Tr}_{Z,N}\, e^{-\beta H}\,,
\ee
where ${\rm Tr}_{Z,N}$ is taken over states of $Z$ protons and $N$ 
neutrons. For small $\zeta_p$ and $\zeta_n$, the quantity $\ln \cz$
can be expanded in a virial series
\be
\ln \cz = {\sum_{Z,N}}^{\prime} D_{Z,N}(\ze_p)^Z (\ze_n)^N ~.
\ee   
Here the prime indicates that the term with $Z=N=0$ is excluded.
The knowledge of the virial coefficients $D_{Z,N}$ gives the partition function
and thence the thermodynamic observables.

In Ref.~\cite{das}, it was shown that all the dynamical information
concerning the microscopic interaction in the grand potential of the 
system is contained in the partition function as two types of terms:
\be
\ln \cz =\ln {\cz}_{part}^{(0)} + \ln {\cz}_{scat}~,
\ee
corresponding to contributions from stable single-particle states of
clusters of different sizes (neutrons and protons included) formed in the 
infinite system and (multiparticle) scattering states, respectively.
The superscript $(0)$ indicates that the clusters behave like an
ideal quantum gas. The first term can further be decomposed
into contributions from ground states and excited states below
nucleon emission threshold. Then,
\be 
\ln {\cz}_{part}^{(0)}=\ln {\cz}_{gr}^{(0)} +\ln {\cz}_{ex}^{(0)} \,,
\ee 
with
\bea
&&\ln {\cz}_{gr}^{(0)}  =  \mp V \sum_{Z,N}g_0 \int\!
\frac{d{\bf p}}{(2\pi)^3}\,\times \nonumber \\ 
 && ~~~~~~~\ln \left( 1\mp \ze_{Z,N}
e^{-\beta(p^2/2Am )} \right) .
\label{eqZ0}
\eea
The upper and lower signs in the above equation correspond to bosons
and fermions, respectively, with $g_0$ as the ground-state spin
degeneracy. For mass numbers $A \le 8$, the $g_0$ values are taken
from experiment. For heavier nuclei, $g_0$ is taken as 1 or 2
depending on whether the nuclei are bosonic or fermionic. The sum in
Eq.\ (\ref{eqZ0}) extends over all possible
fragment species that can be formed; $Z$, $N$, and $A$ refer to the
proton, neutron, and mass number of a species, respectively. Here,
{\bf p} refers to the momentum of the fragment, $m$ is the nucleon
mass, and $V$ is the volume of the system. The {\it effective fugacity}
is given by $\zeta_{Z,N}=e^{\beta (\mu_{Z,N} +B_{Z,N})}$, where the
chemical potential of the fragment is $\mu_{Z,N}=Z \mu_p +N \mu_n $
from the conditions of chemical equilibrium. The quantity $B_{Z,N}$
represents the binding energy of the fragment.

In nuclear matter, only nuclear forces contribute to the binding
energy as the Coulomb interaction is absent; then the fragment size can, 
in principle, be infinite. In supernova matter, the Coulomb interaction 
is operative in determining the binding
energy and the size of the nuclei; the sum in Eq.~(6) is then finite.
The nuclei are embedded in a sea of electrons and the binding energy
gets dressed up; the correction can be estimated using the Wigner-Seitz
approximation \cite{sha}. We work in natural units $\hbar=c=1$.
The integral in Eq.~(6) can be expanded in powers of the effective 
fugacity $\zeta_{Z,N}$ as 
\bea
&&\!\!\!\!\!\!\!\!\ln {\cz}_{gr}^{(0)} =
V\sum_{Z,N}\frac{g_0}{\lm^3_A}\left(\zeta_{Z,N} \pm
\frac{\zeta^2_{Z,N}}{2^{5/2}} +\cdots\right).
\eea
The quantity $\lambda_A= \sqrt{2 \pi/(AmT)}$ is the thermal
wavelength of a species of mass $Am$.

A nucleus in a particular excited state is taken as a distinctly
different species and can be treated in the same footing as the 
ground state. The density of states is quite high in relatively
heavy nuclei and increases nearly exponentially with the square root
of the excitation energy $E^*$. Thus, the contribution of the excited
states of a single nucleus is given as an integral over $E^*$ of the ideal
gas terms weighted with the level density $\omega(A,E^*)$:
\bea 
\ln {\cz}_{ex}^{(0)} &=& \mp V{\sum_{Z,N}}^{\prime}
\int_{E_0}^{E_s} \!\!dE^*~\om(A,E^*) 
\times\nonumber\\
&&\int\!\frac{d{\bf p}}{(2\pi)^3}\, \ln \left( 1\mp
\zeta_{Z,N} e^{-\beta(p^2/2Am +E^*)} \right) .
\eea
The expression for the level density is obtained from the Fermi gas
model of noninteracting nucleons in a nucleus~\cite{boh}
\be
\om (A,E)=\frac{\sqrt{\pi}}{12a^{1/4}}\frac{e^{2\sqrt{aE}}}{E^{5/4}}\,.
\ee
The level density parameter $a$ is taken as $A/8$ MeV$^{-1}$, its
empirical value.
In Eq.~(8), the prime indicates exclusion of light nuclei ($A\leq 8$)
in the sum. The lower limit $E_0$ is dictated by the location of the first
excited state. We take it to be 2 MeV. The upper limit $E_s$ is the 
particle emission threshold taken as 8 MeV. 

The scattering term in Eq.~(4) can be formally written for the system
under consideration as
\be
\ln {\cz}_{scat} =\sum\! \int\!
dE \, \frac{e^{-\beta (E-\mu)}}{2\pi i} \, {\rm Tr}\!
\left({\cal A}S^{-1}(\ep)\frac{\partial}{\partial E}S(\ep)\right)_c,
\ee
where the sum is over all scattering channels, each having its chemical
potential $\mu $ and formed by taking any number of particles from
any of the stable species. The trace is over all the plane wave
states for each of the channels. $S$ is the scattering operator and
${\cal A}$ is the boson symmetrization or fermion antisymmetrization
operator. The subscript $c$, in diagrammatic language,
refers to only the connected parts of the expression in parenthesis.

To recast Eq.~(10) explicitly in the context of nuclear or supernova 
matter, a set of channels with total proton number $Z_t$, neutron number
$N_t$, and mass number $A_t$ will
be considered. All other labels required
to fix a channel are denoted by $\sigma $. Obviously, the total mass and
the chemical potential are independent of $\sigma$, depending only on
$Z_t$ and $N_t$. The nonrelativistic energy in a channel is given by 
\be
E_{Z_t,N_t,\sg}=\frac{P^2_{CM}}{2A_t m} -B_{Z_t,N_t,\sg} +\ep\,,
\ee
where ${\bf P}_{CM}$ is the total center of mass momentum, $\ep$ is
the kinetic energy in the CM frame, and $B_{Z_t,N_t,\sigma }$ is the
sum of the individual binding energies of all the fragments in the
channel. Integrating over ${\bf P}_{CM}$, one then gets
\bea
\ln {\cz}_{scat} &=& V\!\sum_{Z_t,N_t}\frac{e^{\beta \mu_{Z_t,N_t}}}
{\lm^3_{A_t}}\sum_{\sg} e^{\beta B_{Z_t,N_t,\sg}} \times \nonumber \\
&& \int_0^{\infty}\!\!\!d\ep \, \frac{e^{-\beta\ep}}{2\pi i} \,
{\rm Tr}_{Z_t,N_t,\sg}\!\!
\left({\cal A}S^{-1}(\ep)\frac{\partial}{\partial\ep}S(\ep)\right)_c , 
\nonumber \\
\eea
the trace being now restricted to the channel $(Z_t, N_t, \sg)$.
Examination of Eq.~(12) shows that larger binding energies are more
important, because of the factor $e^{\beta B_{Z_t,N_t,\sg}}$. Furthermore,
two-particle channels are expected to be more dominant than the
multiparticle channels with the same $Z_t$ and $N_t$ from binding energy
considerations. The two-particle scattering channels are therefore only
considered.
It becomes convenient to divide the channels into light ones, consisting 
of low-mass particles ($A \le 8$, say) and heavy ones, containing at
least one high-mass particle ($A > 8$), so that we write
\bea
\ln {\cz}_{scat}~=~ \ln {\cz}^L_{scat}+\ln {\cz}^H_{scat} \,.
\eea

The scattering of relatively heavier nuclei is known to be dominated
by a multitude of narrow resonances near the continuum threshold. The
$S$-matrix elements are then approximated by resonances. Each of these
resonances can be treated  \cite{das1,das2} like an ideal gas term.
Then, $\ln{\cz}^H_{scat}$ can be written in the form of
$\ln{\cz}^{(0)}_{ex}$, assuming the resonance level densities to be
the same as those of the excited states given by Eq.~(9). The sum of
the contributions from the excited and the resonance states can then be
written as 
\bea 
&&\ln {\cz}_{ex}^{(0)} + \ln {\cz}^H_{scat}=\mp V{\sum_{Z,N}}^{\prime}
\int_{E_0}^{E_r} \!\!dE^*~\om(A,E^*) 
\times\nonumber\\
&&\int\!\frac{d{\bf p}}{(2\pi)^3}\, \ln \left( 1\mp
\zeta_{Z,N} e^{-\beta(p^2/2Am +E^*)} \right)\,,\\
&&\!\!\!\!\!\!\!\!= V{\sum_{Z,N}}^{\prime} 
\frac{1}{\lm^3_A}\left(f_1\,\zeta_{Z,N} \pm
f_2\,\frac{\zeta^2_{Z,N}}{2^{5/2}} +\cdots\right).
\eea
The integration in Eq.~(14) extends up to $E_r$, the limit of resonance 
domination. The damping of the integral in Eq.~(14) due to the
presence of the
Boltzmann factor limits the contributions to only those from low
energies; we take $E_r$=~12 MeV. The $A$-dependent entities
\bea
f_n(A)~=~\int_{E_0}^{E_r}dE^* \om (A,E^*)e^{-n\beta E^*},
\eea
with $n =1,2,\cdots $, decrease steadily with increasing $n$, so that
the series converges quite fast.
For the evaluation of $\ln {\cz}^L_{scat}$ (i.e., the contribution of
light particles to $\ln {\cz}_{scat}$), only the scattering
channels $NN, Nt, N {\rm He}^{3}, N\alpha $,  and $\alpha \alpha $
are considered, where $N$ and $t$ refer to the nucleon and
the triton, respectively. Then,
\be
\ln {\cz}_{scat}^L~=~ \ln {\cz}_{NN}+\ln {\cz}_{Nt}
+\ln {\cz}_{N{\rm He}^3}
+\ln {\cz}_{N\alpha}+\ln {\cz}_{\alpha \alpha}.
\ee
Each of the terms in Eq.~(17) can be expanded in the respective virial 
coefficients. We consider the expansion up to the second-order
coefficients which are written as energy integrals in terms of the
relevant phase shifts.

In summary, the grand partition function for the interacting nuclear
system is given as 
\bea
\ln {\cz}=\ln {\cz}_{gr}^{(0)}+\left (\ln {\cz}_{ex}^{(0)}+
\ln {\cz}^H_{scat} \right )+\ln {\cz}^L_{scat} \,.
\eea
Once the partition function is known, the chosen observables
can be calculated. Expressions for them in some detail are
given in the following subsections.

\subsection {Equation of state} 

The expression for $\ln {\cz}$ is given by
\bea
&&\ln {\cz}=V\{\frac{2}{\lambda_N^3}  [ \zeta_n +\zeta_p+
\frac{b_{nn}}{2}\zeta_n^2+\frac{b_{pp}}{2}\zeta_p^2 
+\frac{1}{2}b_{np}\zeta_n\zeta_p ] \nonumber \\
&& +\frac{2}{\lambda_t^3}[ \zeta_t+2\zeta_t (b_{pt}\zeta_p+b_{nt}\zeta_n)]
\nonumber \\
&& +\frac{2}{\lambda_h^3}[ \zeta_h+2\zeta_h (b_{ph}\zeta_p+b_{nh}\zeta_n)]
\nonumber \\
&&+\frac{1}{\lambda_\alpha^3}[\zeta_\alpha+b_{\alpha\alpha}\zeta_\alpha^2
+b_{\alpha n}\zeta_\alpha (\zeta_n +\zeta_p)]\} +\ln{\cz}_{Rest} \,.
\eea
In Eq.~(19), the subscripts $N,t,h$, and $\alpha$ refer 
to the nucleon, triton, He$^{3}$,
and He$^{4}$, respectively. The coefficients $b$ are the virial coefficients.
In the limit of isospin symmetry, we take $b_{nn} = b_{pp}$. Due to lack
of $p\,t$ scattering data, we assume $b_{pt} \simeq b_{nh}$. These
virial coefficients can be written in terms of experimentally known
phase shifts. As an example, we write below explicit expressions for
$b_{nn}$ and $b_{np}$:
\bea
b_{nn}=-\frac{1}{2^{3/2}}+ \frac{\sqrt{2}}{\pi T}\int_0^\infty dE~
\delta_{nn}^{tot}(E)~e^{-\beta E/2},
\eea
and
\bea
b_{np}=b_{np}^0+b_d \,,
\eea
with
\bea
b_d=6\sqrt{2}~e^{B_d/T}
\eea
and
\bea
b_{np}^0=-6\sqrt{2}+\frac{\sqrt {2}}{\pi T}
\int_0^\infty \delta_{np}^{tot}(E)~e^{-\beta E/2}.
\eea
In Eq.~(21), the term $b_{np}^0$ corresponds to the non-resonance
$n-p$ scattering contribution;
the term $b_d$ corresponds to the resonance contribution coming from
the bound state of the deuteron with binding energy $B_d$. The energy
$E$ is measured in the laboratory frame. The expression for the total
phase shift is given as
\bea
\delta_{NN}^{tot}=\sum_{LSJ}(2J+1) \{\delta^{2S+1}_{L_{J}}(I=0)
+\delta ^{2S+1}_{L_{J}}(I=1) \} .
\eea
The contributing partial waves are determined by the isospin $I$
with the requirement of the antisymmetry on the total wave function 
of the $NN$ system. 

The last term in Eq.~(19) is the sum of the contributions
from the rest of the species ($A>4$) and is given by
\bea
\ln {\cz}_{Rest}=V\sum_{i}\frac{\zeta_i}{\lambda_i^3} \!
\left( g_0+\int_{E_0}^{E_r}\!\! \omega (E^*)e^{-E^*/T}dE^* \right) \! . 
\eea
As already stated, the second term in Eq.~(25) contributes only for 
$A>8$. The pressure can be evaluated from 
\bea
P=T \ln {\cz}/V \,.
\eea
The number density $\rho_i$ of the $i$-th fragment species is calculated
from 
\bea
\rho_i=\zeta_i \left (\frac{\partial}{\partial \zeta_i} 
\frac{\ln {\cz}}{V} \right )_{V,T}.
\eea
The total neutron, proton and baryon density in the system can be
obtained from
\bea
\rho_n^B=\sum_{i}N_i\rho_i, \nonumber  \\
\rho_p^B=\sum_{i}Z_i\rho_i, \nonumber  \\
\rho^B=\sum_{i}A_i\rho_i. 
\eea

From the Gibbs-Duhem relation, the free energy density is 
\bea
{\cal F}=-P+\sum_i\mu_i\rho_i.
\eea
The entropy density ${\cal S}$ is calculated from
\bea
{\cal S}=\left(\frac{\partial P}{\partial T}\right)_\mu,
\eea
which then yields the total energy density as
\bea
{\cal E}_{tot}={\cal F} + T{\cal S} \,.
\eea
The detailed expression for the energy density is
\bea
&&{\cal E}_{tot}=\frac{3}{2}T\sum_i\rho_i+\sum_{i\in H}\frac{\zeta_i}
{\lambda_i^3}\int_{E_0}^{E_r}\omega (E^*)E^*e^{-E^*/T}dE^* \nonumber  \\ 
&&-\sum_i\rho_i B_i  
 -\frac{3}{2}T\{\frac{1}{\lambda_N^3}(b_{nn}\zeta_n^2+b_{nn}\zeta_p^2
+b_{np}^0\zeta_n\zeta_p)  \nonumber \\
&& +\frac{4}{\lambda_t^3}\zeta_t(b_{nt}\zeta_n+b_{pt}\zeta_p)
+\frac{4}{\lambda_h^3}\zeta_h(b_{nh}\zeta_n+b_{ph}\zeta_p) \nonumber \\
&&+\frac{1}{\lambda_{\alpha}^3}[b_{\alpha n}\zeta_\alpha (\zeta_n+
\zeta_p)+b_{\alpha\alpha}\zeta_{\alpha}^2]\}
+\frac{T^2}{\lambda_N^3}\{b_{np}^{0\prime }\zeta_n\zeta_p \nonumber \\
&&+b_{nn}^\prime (\zeta_n^2+\zeta_p^2)\}  
+\frac{4T^2}{\lambda_t^3}\zeta_t(b_{nt}^{\prime}\zeta_n+b_{pt}^{\prime}
\zeta_p) \nonumber \\
&&+\frac{4T^2}{\lambda_h^3}\zeta_h(b_{nh}^{\prime}\zeta_n
+b_{ph}^{\prime}\zeta_p) \nonumber \\
&& +\frac{T^2}{\lambda_{\alpha}^3}\{b_{\alpha\alpha}^{\prime}
\zeta_\alpha ^2+b_{\alpha n}^{\prime }\zeta_\alpha (\zeta_n+\zeta_p)\}
\,,
\eea
where $\sum_{i\in H}$ denotes that the sum runs over the channels of
heavy particles.
In Eq.~(32), the first term is identified with ${\cal E}_{CM}$, the
kinetic energy density associated with the center of mass of the fragments.
The second term refers to ${\cal E}^*$, the sum 
of the  densities of the thermal and
resonance excitation energies of the fragments, and the third term
coming from the negative of the sum of the fragment
binding energies $B_i$
is denoted as ${\cal E}_{BE}$.
 The rest is designated as ${\cal E}_I^{LL}$,
the sum of the contributions coming from the interactions between
different pairs of light fragments. The primes on the virial coefficients
denote their temperature derivatives. They are part of the entropy
contributions. Thus, Eq.~(32) can be rewritten as
\bea
{\cal E}_{tot}={\cal E}_{CM}+{\cal E}^*+{\cal E}_{BE}+{\cal E}_I^{LL}.
\eea

\subsection{The single-particle and isovector potentials
 and the symmetry coefficients}

The single-particle potentials for neutrons or protons in a
nuclear medium are conventionally defined as \cite{boh}
\bea
V_{\tau}=V_0+\tau V_1 X.
\eea
Here, $V_0=(V_n+V_p)/2$ is the isoscalar potential, 
$V_1$ is the measure of
the isovector potential, and $\tau = \pm \frac{1}{2}$ for neutrons
or protons. 
The quantity $X = (\rho_n^B-\rho_p^B)/\rho^B $ is the
asymmetry parameter,
with $\rho_{\tau}^B$ as the total neutron or proton number densities
in the system, and $\rho^B= \rho_{\tau}^B+\rho_{-\tau}^B $ is the
total nucleon density. The isovector potential $ V_{isov}$ is then 
\bea
V_{isov}=V_n-V_p=V_1 X.
\eea

In the nucleonic medium, the single-particle potentials are obtained
as 
\bea
V_{\tau}=\left ( \frac{\partial {\cal E}_I}{\partial \rho_{\tau}^B}
\right )_{\rho_{-\tau}^B},
\eea
where ${\cal E}_I$ is the interaction energy density. In homogeneous
nuclear matter,  it is calculated
from the effective interaction. In clusterized nuclear matter, 
we employ Eq.~(36) 
for the definition of the {\it effective}
single-nucleon potential. The calculation of ${\cal E}_I$
is, however, not straightforward. We have adopted the following procedure
to take counts of the contributions to the interaction energy density.

In Eq.~(33), the first term is purely kinetic and the last term comes
solely from the interactions. The binding and the excitation
energies, however, are admixtures of both kinetic and interaction
contributions. The binding term ${\cal E}_{BE}$ can be split as
\bea
{\cal E}_{BE}=-\sum_i \rho_i B_i=-\sum_i\rho_i (B_i^K+B_i^I).
\eea
In Eq.~(37), the two terms are the kinetic and the interaction parts of the
binding energy in a fragment, respectively. The kinetic terms are 
estimated with the choice of the extended Hu$^\prime $lthane wave function 
for the deuteron \cite{adl} and a Gaussian for $t$, He$^{3}$, and
$\alpha $ particles \cite{sch}.
The kinetic terms for these light nuclei are very small compared
to the total binding term of the system, hence, 
even a change by 50$\%$ in the estimated values of the kinetic terms
in the light nuclei
does not change the values of the calculated observables noticeably.
For heavier systems, the Fermi gas approximation is employed. Since 
the binding energies are known experimentally, the interaction
contribution $B_i^I$ can then be obtained. Estimation of the interaction
contribution ${\cal E}_I^*$ from the excitation part ${\cal E}^*$ is not
easy. However, compared to the binding term $\sum_i \rho_i B_i^I $,
it is quite small. It is checked that changes in the 
 single-particle and isovector potentials  are 
insignificant
with ${\cal E}_I^*$=0 or ${\cal E}_I^*={\cal E}^*$. The calculations 
reported correspond to ${\cal E}_I^*= \frac{1}{2} {\cal E}^*$. The 
interaction energy density then takes the form,
\bea
{\cal E}_I=\frac{1}{2}{\cal E}^*-\sum_i\rho_i(B_i-B_i^K)+{\cal E}_I^{LL}, 
\eea
where we recall that ${\cal E}_I^{LL}$ refers to the contribution from
the interactions between different pairs of light fragments. Using
Eqs.~(27), (36), and (38), the effective 
 single-particle and isovector potentials in clusterized matter
can then be calculated. The single-particle potentials are called
effective in the sense that they represent the average of the
interaction of a single nucleon with all other free nucleons and
bound nucleons in clusters of different sizes, the clusters being
present in different proportions in the inhomogeneous system.

The total symmetry energy coefficient $C_s^{tot}$ is defined through
\bea
e(X)=e(X=0)+C_s^{tot} X^2,
\eea
where $e={\cal E}_{tot}/\rho^B$ 
is the energy per nucleon. The coefficient $C_s^{tot}$
can be split into interaction and kinetic terms as 
\bea
C_s^{tot}=C_s^I+C_s^K,
\eea
corresponding to
\bea
e_I(X) &=& e_I(X=0)+C_s^I X^2, \nonumber \\
e_K(X) &=& e_K(X=0)+C_s^K X^2.
\eea
In nuclear matter, the isovector potential is related to the
interaction energy per nucleon $e_I(X)$ as 
$V_{isov}=2 (\partial e_I(X)/\partial X)_{\rho^B}$. For homogeneous
matter, $e_I(X)$ is linear in $X^2$, then
\bea
V_{isov}=4 C_s^I X.
\eea
For clusterized matter, this relation is, however, only approximate
because $e_I(X)$ (as well as the energy per nucleon $e(X)$ \cite{de})
is seen to be not fully linear in $X^2$.

The incompressibility $K(X)$ of a nuclear system is calculated from
\bea
K(X)=9 \frac{dP}{d\rho^B},
\eea
where $P$, given by Eq.~(26), is the pressure of the system 
with asymmetry $X$. It can be written as
\bea
K(X)=K(X=0)+K_s X^2,
\eea
where the coefficient $K_s$ is the symmetry incompressibility.

\section{Results and Discussions}

We have calculated the equation of state (EOS),  
the single-particle and isovector potentials, symmetry 
energy coefficients and the symmetry incompressibility of dilute nuclear
and supernova matter at different temperatures and densities. Results for
nuclear matter are presented in subsection A and those for supernova matter
are given in subsection B. Calculations have been restricted up to a baryon
density $\rho^B$ =~0.02 fm$^{-3}$. At relatively high density, the asymptotic
wave function may not have a precise meaning and then expressions of the
partition function in terms of $S$-matrix elements may not be very meaningful.

The virial coefficients related to $NN, N\alpha$, and $\alpha\alpha$
scatterings alongwith their temperature derivatives are taken from
Ref.~\cite{hor}. The same for $Nt$ and $Nh$ scatterings are obtained
from Ref.~\cite{con}. In doing so, appropriate care has been taken
for the slightly different choices of factors in the virial expansion
of the partition function.

\subsection{Nuclear matter}

In Fig.~1, the pressure $P$ is displayed as function of baryon density
for symmetric ($X$=0.0) and asymmetric ($X$=0.3) nuclear matter. 
The upper panel corresponds to a temperature $T$=~4 MeV, the
lower panel corresponds to $T$=~8 MeV. The results calculated in the
$S$-matrix approach (SM) are compared with those in a mean-field (MF)
model. The MF calculations are performed with the SkM$^*$ interaction. 
At very low densities, $P$ is nearly model independent ($P\simeq \rho T$);
at a little higher density, the difference between the two models is
apparent from the figure. In the MF calculation, with isothermal compression,
the dilute system enters the unphysical 
region beyond a certain density 
which increases with temperature. 
In the SM approach, with compression, the unphysical
behavior does not arise because of many-body correlations (condensation).
For symmetric matter, at low temperature, the pressure levels off at
very low densities as shown by the full line in the upper panel signaling
a behavior like a first-order phase transition. 
It points out a phase coexistence between light and heavy clusters. 
For asymmetric matter, the rise in pressure is linear in density at constant
temperature for very dilute systems. There is
a break from this linearity at a certain density (we refer to this density as
the condensation density); the
 pressure thereafter rises monotonically
behaving like a second-order phase transition.
At the higher temperature
$T$= 8 MeV, the said transition occurs at a  much higher density 
which is beyond the density we consider.
The increase in free neutron multiplicity with density is mainly
responsible for this monotonic rise. For both symmetric and asymmetric
systems, the chemical equilibrium conditions coupled with the conservation
of the baryon number and isospin governs this behavior.

Since there is no Coulomb in the binding energy of the fragments formed
in nuclear matter, the sum in Eq.~(3) runs up to infinity in principle;
in practice, one takes a finite sum for calculational facilitation. The
calculations here have been performed with a maximum fragment mass 
$A_{max}$=1000. The results are not very sensitive to further increase
in the maximum mass \cite{de}. The binding energies of these nuclei
are obtained from the liquid-drop type mass formula \cite{dan} with
Coulomb switched off.

The effective single-particle potentials $V_n$ and $V_p$ for neutron
and proton  are shown in Fig.~2
as a function of density at temperatures $T$=4 and 8 MeV 
for symmetric and asymmetric ($X$=0.3) nuclear matter. The
isovector potential $V_n-V_p$ is also shown in the bottom panels
for asymmetric matter at the same temperatures.
Calculations have been performed
in both the SM and MF approaches. In the mean-field model, the single-nucleon
potentials decrease with increasing density with a monotonic increase
of the isovector potential. They are nearly independent of temperature.
In the $S$-matrix approach, the single-nucleon potentials show a subtle
behavior with density.
For symmetric matter, $V_n$ (=$V_p$), at lower temperature decreases 
sharply upto the condensation density, beyond which it remains
practically constant. Beyond this density, the fragment composition of
matter scales nearly with  density and the constancy of the
single-particle potential is a reflection of that. The value of this
constant effective single-particle potential is $\sim -34$ MeV in
contrast to that of $\sim -60$ MeV for saturated uniform nuclear
matter. This is so because in calculating $\delta {\cal E}_I/\delta \rho $
for uniform matter, one has to count the change in nucleon density,
but in clusterized matter, the internal nucleonic density of fragments
remains unaltered, only their number density changes. At lower temperature,
for asymmetric nuclear matter, $V_n$ goes through a sharp minimum
around the condensation density. Beyond this density, it passes
through a maximum and then decreases slowly; $V_p$ on the other hand
behaves more like that of symmetric nuclear matter, but it is
deeper. The above qualitatively different behavior 
of $V_n$ and $V_p$ comes from
the presence of free neutrons in asymmetric nuclear matter.  
The isovector potential passes through a maximum and then
decreases slowly with density. As in mean-field, the isovector potential
is seen to be nearly proportional to the asymmetry $X$. The different
nature of the single-particle potentials at higher temperature
as seen for $T$=8 MeV is a manifestation of the dilution of condensation
effects.

The symmetry energy coefficients, as a function of baryon 
density, are displayed in the left and right panels of Fig.~3 at temperatures
$T=$4 and 8 MeV, respectively.
In the MF approach,
the total symmetry coefficient $C_s^{tot}$ alongwith the kinetic
and interaction components $C_s^K$ and $C_s^I$ (shown as dot-dash lines) 
increase linearly
with density and are seen to be practically independent of temperature.
The symmetry energy $e_{sym}(X)=e(X)-e(X=0)$
in the SM approach, as opposed to that
in the MF model, is found to be anharmonic in $X$, particularly
at lower temperatures; this was already noted earlier \cite{de}.
This induces an asymmetry dependence in the symmetry coefficients
defined through Eqs.~(39) and (41). To avoid this, we therefore,
take the definition \cite{hor} 
\bea
C_s=\frac{1}{2}\left(\frac {\partial^2e(X)}{\partial X^2}\right )_{X=0}.
\eea
This applies to
$C_s^I, C_s^K$ and $C_s^{tot}$ with appropriate
choice of the energy components. The symmetry coefficients, so
defined are shown as full lines in the figure. The symmetry coefficients
thus obtained are seen to be very different from those obtained in the
MF model. The magnitudes of the coefficients are much larger, the
kinetic component is always negative 
and there is a marked 
dependence on temperature. The negative symmetry kinetic energy
$C_s^K X^2$ ($=e_K(X)-e_K(X=0)$) looks antiintuitive. For clusterized
nuclear matter, it, however, can be understood from the fact that
for symmetric matter, clusterization is more favored leading to
larger internal kinetic energy compared to asymmetric matter where
there are more free neutrons and have lesser total internal
kinetic energy.

The anharmonicity of the symmetry energy $e_{sym}(X)$
(calculated in the SM approach through Eq.~(32)) in the asymmetry
parameter  is portrayed in Fig.~4. In the figure the evolution
of $D_{SE}$, a measure of this anharmonicity, is shown as a function
of baryon density $\rho^B$ at different $X$ at temperatures $T=$4
(full line) and 8 MeV (dashed line), respectively. The anharmonicity
parameter $D_{SE}$ is defined as
\bea
D_{SE}=\frac {C_s^{tot}X^2-e_{sym}(X)}{C_s^{tot}X^2}\times 100~.
\eea
It represents the percentage deviation of $e_{sym}(X)$ from
$C_s^{tot}X^2$ where $C_s^{tot}$ is 
defined through Eq.~(45). As expected, anharmonicity
increases with asymmetry. 
On the other hand, it decreases with increasing
temperature because of the dissolution of clusters with heating.
  
 In Fig.~5, the incompressibility coefficient $K$ of symmetric ($X$=0.0) and 
asymmetric ($X$=0.3) nuclear matter are presented in the
left panels as a function of 
density at $T=4$ and 8 MeV. At very low density, the incompressibility
in both the SM and MF approaches is nearly the same. Increasing density
and subsequent condensation renders the system more compressible in the
SM approach. The somewhat different behavior of the incompressibility 
for symmetric and asymmetric matter at lower temperature in the $S$-matrix 
approach can be easily understood from the 
different functional dependence of pressure with density for
symmetric and asymmetric systems as shown in Fig.~1. In
the MF approach, on isothermal compression,
the system enters the unphysical region beyond a certain density 
where the incompressibility
becomes negative. This is not shown in the figure.
From the right panels of this figure,  it is seen that
in the physical region, at a fixed density, the incompressibility $K$
decreases almost linearly on isochoric cooling. In the SM
approach, on isochoric cooling, there is a linear decrease in $K$,
 however, a sudden fall in incompressibility at a certain temperature is
noticed. This is related to the onset of condensation where the
fragment multiplicity suddenly drops \cite{de1}.

 The symmetry incompressibility $K_s$ defined in Eq.~(44) is compared
in the SM and MF models in Fig.~6 at temperatures $T$=4 and 8 MeV. 
As opposed to the MF model where $K_s$ increases monotonically in the
density region we explore, the same in the SM approach has a peaked
structure which is prominent at lower temperature. Here, the
symmetry incompressibility is seen to be weakly 
dependent on the asymmetry of the system 
while in the MF model it is asymmetry independent.
The arrows in the figure indicate the density above which the system enters
the unphysical region in the mean-field model.

    In Fig.~7, the symmetry incompressibility $K_s$ is shown
as a function of temperature at baryon densities $\rho^B$=
0.001 and 0.01 fm$^{-3}$.  In contrast
to the mean-field results where $K_s$ is practically independent of
temperature, in the $S$-matrix approach, $K_s$ is sharply peaked
around the condensation region. The asymmetry dependence of $K_s$
in this approach washes out with increasing temperature. The arrow
in the lower panel of this figure 
marks the temperature below which the system is in
the unphysical region in the MF model.

\subsection{Supernova matter}

Neutrons, protons and electrons are assumed to be the basic
constituents of the supernova matter we are dealing with. The 
matter is also assumed to be in $\beta$-equilibrium. For charge
neutrality, the electron number density $\rho_e$ = $\rho^B_p$, the 
proton number density. This $\beta$-equilibrated matter is treated 
both in the mean-field (MFB) and the $S$-matrix (SMB) approaches.
The proton fraction in the system depends on 
its baryon density and temperature.
In supernova matter,
in the SMB approach, the fragments have coulomb contribution in the binding
energy; so, in the sum in Eq.~(3), only nuclei within the drip lines
are considered. For this purpose, all the isotopes (around 9000
in number) and their binding energies are taken from Ref.~\cite{mye}.
The effect of electron environment is taken into account through the
dressing-up of the fragment binding energies in the Wigner-Seitz
approximation \cite{sha,ish}. For a given electron density $\rho_e$,
the effective binding energy of the $i$-th fragment species 
with $A_i$ and $Z_i$ as its mass and atomic number is given by,
\bea
B_i(\rho_e)~=~B_i(0)+\Delta B_i,
\eea
where 
\bea
\Delta B_i=\frac{3}{5}\frac{Z_i^2e^2}{R_{0i}}  \left(\frac{3}{2}\eta_i
-\frac{1}{2}\eta_i^3  \right)
\eea
with 
\bea
\eta_i~=~ \left(\frac{\rho_e}{\rho_0}\frac{A_i}{Z_i}\right)^{1/3}.
\eea
 In the above equations, $\rho_0$ is the saturation density of normal
nuclear matter, $R_{0i}$=~1.16 $A_i^{1/3}$ fm is the  radius
of the fragment of mass $A_i$ and charge $Z_i$. For proton, $R_{0i}$
is taken as 0.8 fm. The increase in the effective binding energy
acts to enhance the formation of heavier nuclei.

The electron fractions $Y_e$ (=$\rho_e/\rho^B$) for the
supernova matter are compared in the SMB and MFB models as a function
of baryon density $\rho^B$ at temperature $T=$4 and 8 MeV in Fig.~8.
The two results follow nearly the same trend (increasingly neutron-rich
with increasing baryon density); the SMB model, however, 
predicts somewhat less 
neutron-rich system indicating an effectively lower $e^-$-capture rate
which may be significant enough to influence supernova dynamics \cite{bet},
particularly at lower temperature.
In Fig.~9, the calculated baryonic pressure from the
two models at the aforesaid temperatures are shown as a function of density.
In contrast to nuclear matter, the baryonic pressure for supernova
matter in the two models are seen to be nearly the same. 
This is due to the fact that in supernova matter,
to maintain $\beta$-equilibrium, the asymmetry is very high, so
much of the matter is in the free neutron state that effectively 
controls the behavior of the pressure. 

The variation of the symmetry energy coefficients ($C_s^I, C_s^K$, and
$C_s^{tot}$) of supernova matter as a function of density at temperatures
$T=$4 and 8 MeV are shown in Fig.~10.
These coefficients are evaluated using Eqs.~(39) and (41) at the relevant
asymmetries. 
 The MFB results are the same as those 
of nuclear matter as shown in Fig.~3 as they are asymmetry independent.
In the $S$-matrix approach to supernova matter, the symmetry energy
coefficients have the same trend with density as in nuclear matter,
however, the values get much reduced. 
The presence of Coulomb in the fragment binding energy hinders the formation
of heavier fragments and hence this reduction.

The baryonic incompressibility of supernova matter is shown in Fig.~11
as a function of the baryonic density in the SMB and MFB models at $T=$4
and 8 MeV. The results in the two models are qualitatively the same 
at both the temperatures and are different from those obtained for
nuclear matter shown in Fig.~5. The calculated incompressibility
for supernova matter is seen to be significantly larger 
to that in nuclear matter except
at very low density where both nuclear and supernova matter are
mostly composed of free nucleons.

 The baryonic symmetry
incompressibilities $K_s$ for supernova matter in the two models
at the aforesaid temperatures are displayed in Fig.~12. In the
mean-field model, $K_s$ is independent of asymmetry, hence the
MFB results are the same as the MF results as shown in Fig.~6. 
In the $S$-matrix approach, the symmetry incompressibility for
supernova matter at a temperature of 4 MeV is much reduced compared to 
that in nuclear matter at
lower density; at higher temperature, however, the corresponding
two results are close due to absence of formation of heavier fragments
within the density range studied.
The differences between the incompressibilities obtained for
nuclear matter and supernova matter arise because calculations
in nuclear matter refer to a fixed asymmetry which is taken to be
relatively low, but the calculations for supernova matter, with
changing density, pertain to varying asymmetries which turn out
to be quite high. Furthermore, the presence of the Coulomb
contribution in the fragment binding energies precludes the formation
of very heavy fragments, thereby increasing the pressure.

 \section{Concluding remarks}

The low-density nuclear or supernova matter is not homogeneous,
it is clusterized at a finite temperature. In this paper, we have 
investigated some structural properties of this low density matter
in the the $S$-matrix framework. Special emphasis has been given 
to understand the effects brought out by the isospin structure of
the nuclear force. In particular, we have considered its effects 
on the energy, incompressibility and the single-nucleon potentials.
The effects are more clearly manifest in the symmetry energy
coefficients, symmetry incompressibility and isovector potential.

In the $S$-matrix approach, symmetric or asymmetric nuclear matter
condenses on isothermal compression thereby avoiding the unphysical
region brought out in the mean-field model. Condensation eases
pressure, the incompressibility is thus generally lower. Condensation
has a marked effect on the symmetry incompressibility $K_s$. Around
the condensation density, $K_s$ shows a peaked structure,
particularly at lower temperature, in contrast to its smooth
monotonic increase with density in the mean-field model. In
the supernova matter, the isotherms, so also the incompressibility
in the two models are, however, not very different; the rapid
neutronization in the $\beta$-equilibrated matter with increasing
density is mostly responsible for  this change.
A remarkable feature of the symmetry energy in the
$S$-matrix approach  for nuclear matter is that as opposed to
results from mean-field, the symmetry energy is nonlinear in 
$X^2$. 
The symmetry coefficients are also larger in magnitude in the SM model and
that the kinetic components are, contrary to expectations, negative
The same trends follow in supernova matter.

In the $S$-matrix framework, the calculated properties like
the equation of state, the total symmetry energy coefficient
and the incompressibilities of nuclear or supernova matter 
do not depend on any particular choice of the nuclear force,
they are model-independent. They can be directly
connected to the experimentally measured  phase shifts and
the binding energies of the fragments constituting the matter.
The $e^-$-capture envisaged through the chemical equilibrium
conditions in supernova matter is similarly independent of
any model for nuclear interactions. At low temperatures,
in a selective density range, the $S$-matrix model predicts
a somewhat higher electron fraction in the system compared
to the mean-field calculations indicating lower $e^-$-capture
rate. These may, however, play a significant role in supernova
dynamics.

\begin{acknowledgments}
S.K.S. and J.N.D. acknowledge support of DST, Government of India.
M.C. and X.V. partially supported by the Consolider Ingenio
2010 Programme CPAN CSD2007-00042 and grants FIS2008-01661 from MEC
and FEDER and 2005SGR-00343 from Generalitat de Catalunya.
\end{acknowledgments}

\newpage

\begin{figure}
\includegraphics[width=1.0\columnwidth,angle=0,clip=true]{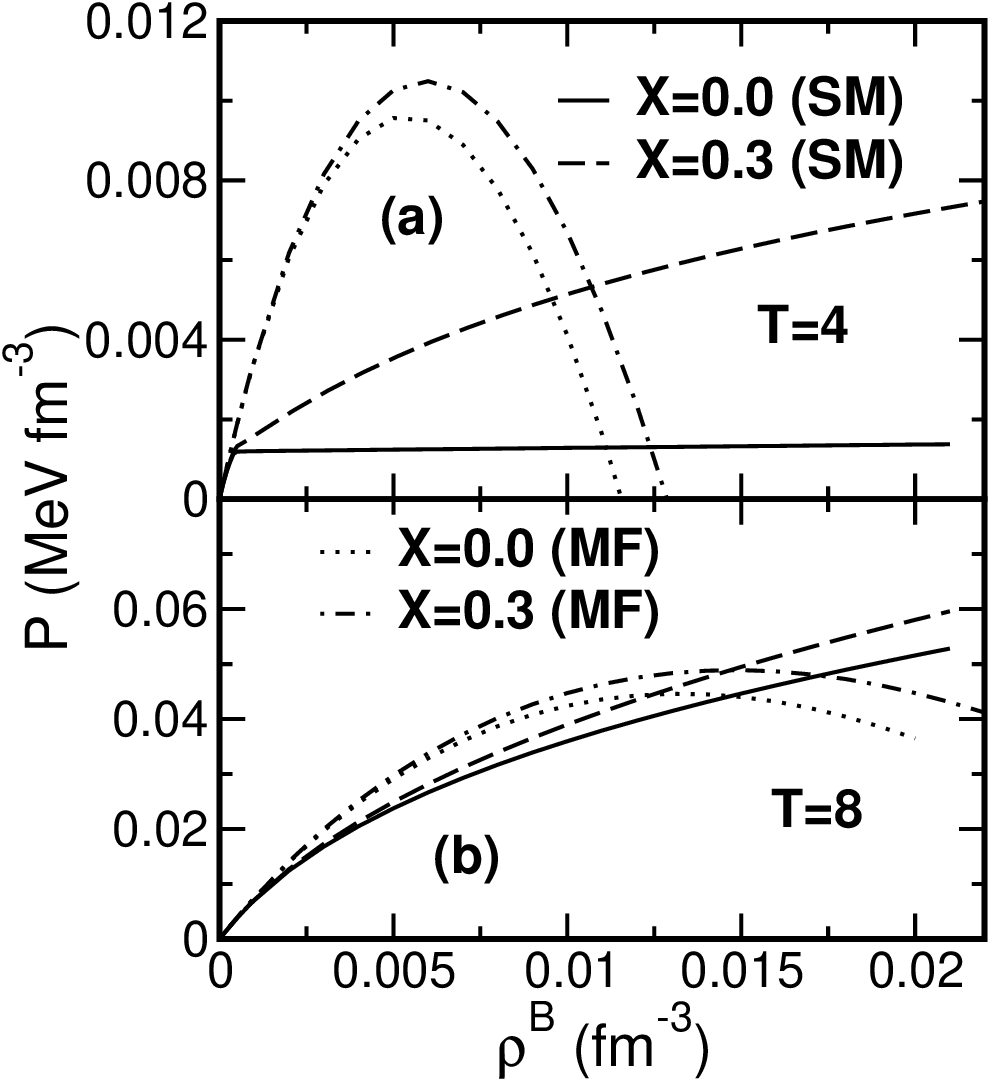}
\caption{ The isotherms for nuclear matter at $T=$4 MeV (upper
panel) and $T=$8 MeV (lower panel) for symmetric ($X=$0.0) and
asymmetric (X=0.3) nuclear matter in the $S$-matrix (SM) and mean-field 
(MF) models.}
\end{figure}

\begin{figure}
\includegraphics[width=1.2\columnwidth,angle=0,clip=true]{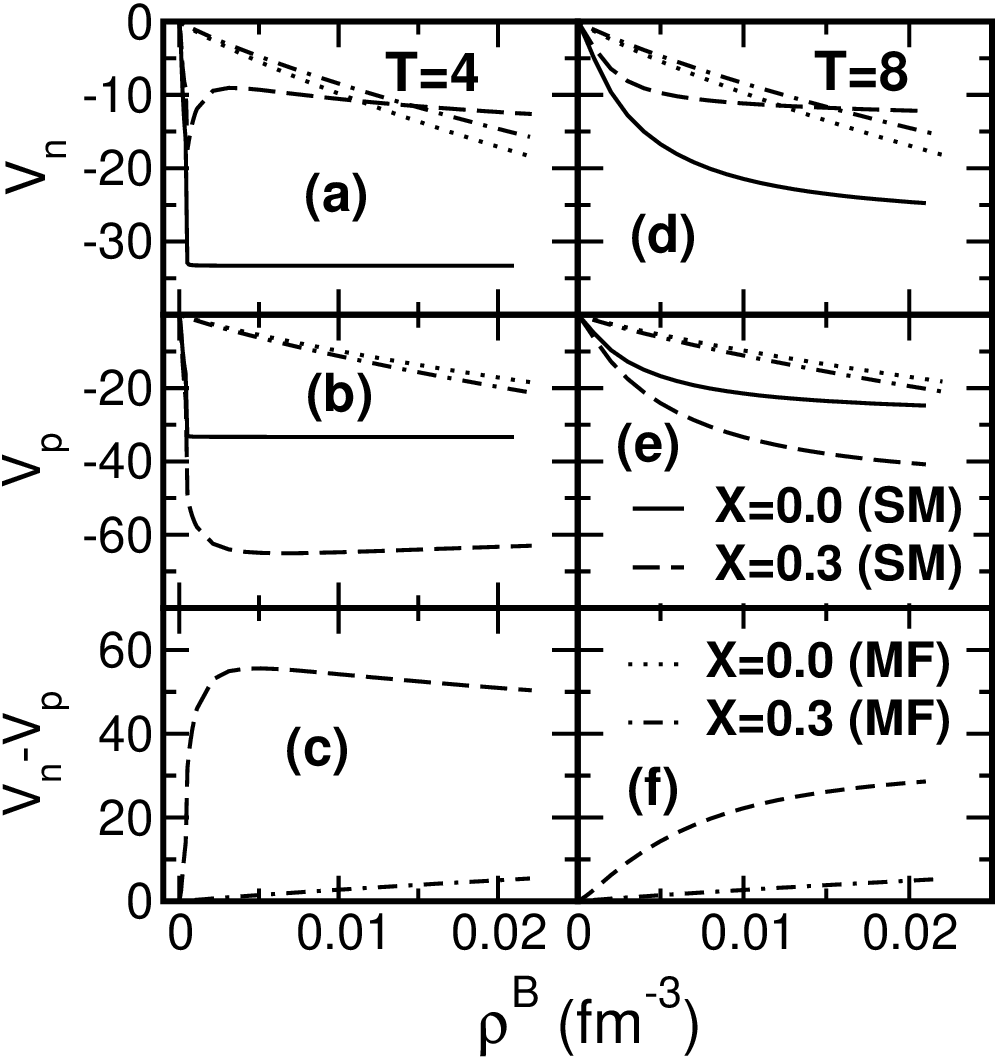}
\caption{ The single-nucleon ($V_n$ and $V_p$) and isovector
($V_n-V_p$) potentials in MeV shown at $T=$4 MeV (left panels) and
$T=$8 MeV (right panels) for symmetric ($X=0.0$) and 
asymmetric ($X=0.3$) nuclear
matter in the SM and MF models as a function of the baryon density.}
\end{figure}

\begin{figure}
\includegraphics[width=1.2\columnwidth,angle=0,clip=true]{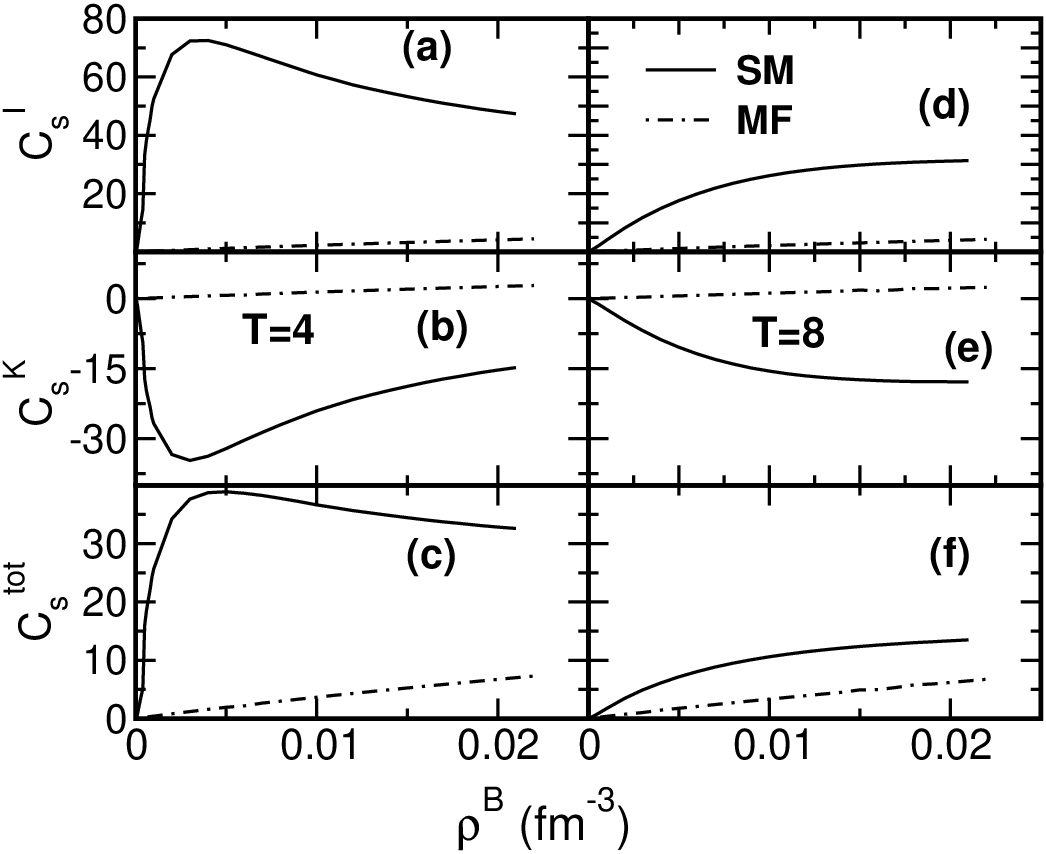}
\caption{ The interaction, kinetic, and total symmetry energy
coefficients ($C_s^I, C_s^K$, and $C_s^{tot}$, respectively,
all in MeV) for
nuclear matter shown as a function of the baryon density at $T=$4 MeV
(left panels) and $T=$8 MeV (right panels) 
in the SM and MF models.} 
\end{figure}

\begin{figure}
\includegraphics[width=1.2\columnwidth,angle=0,clip=true]{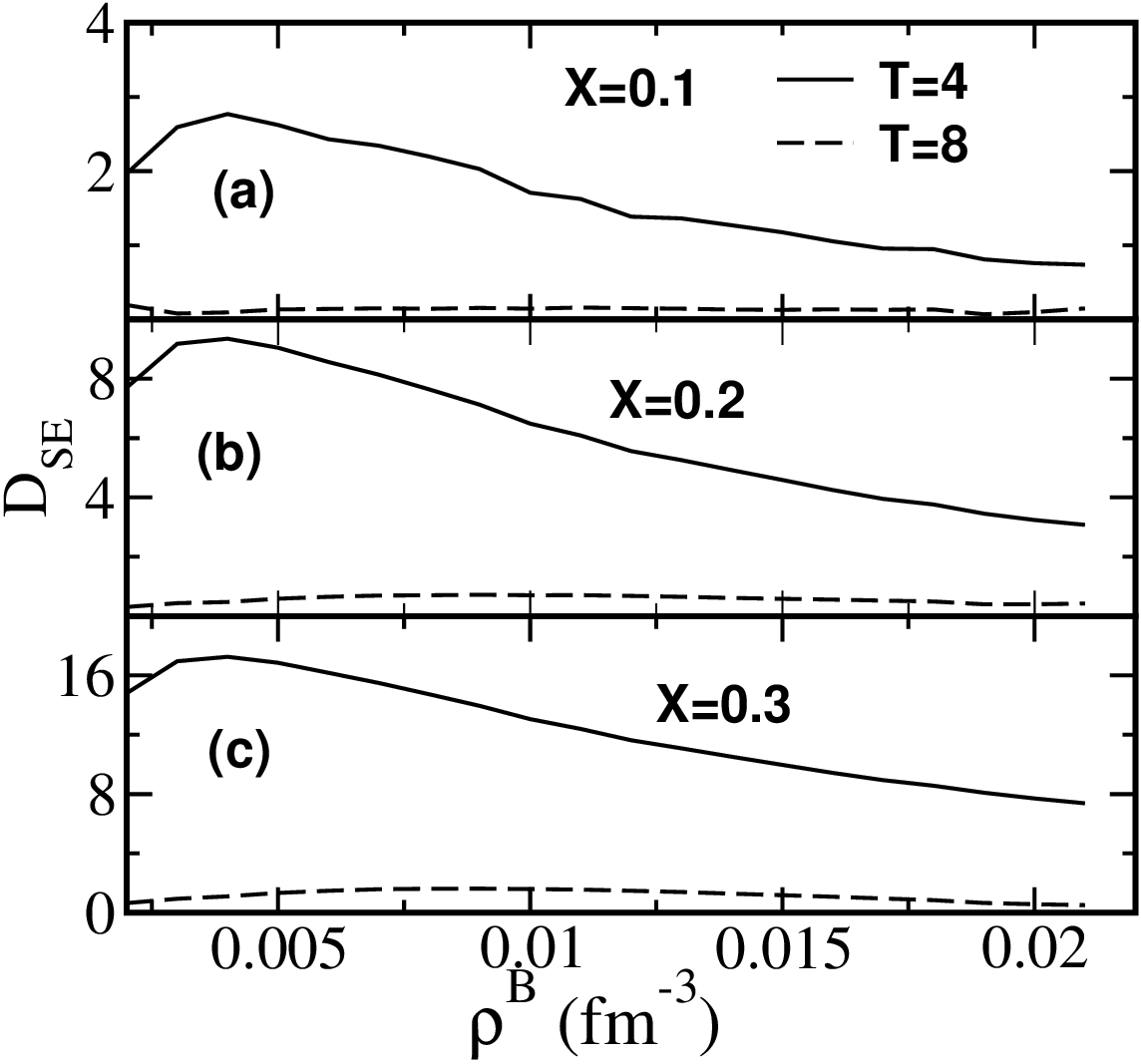}
\caption{The anharmonicity $D_{SE}$ of the symmetry energy 
(see text) plotted
as a function of the baryon density at $T=$4 and 8 MeV for three
different asymmetries.}  
\end{figure}

\begin{figure}
\includegraphics[width=1.2\columnwidth,angle=0,clip=true]{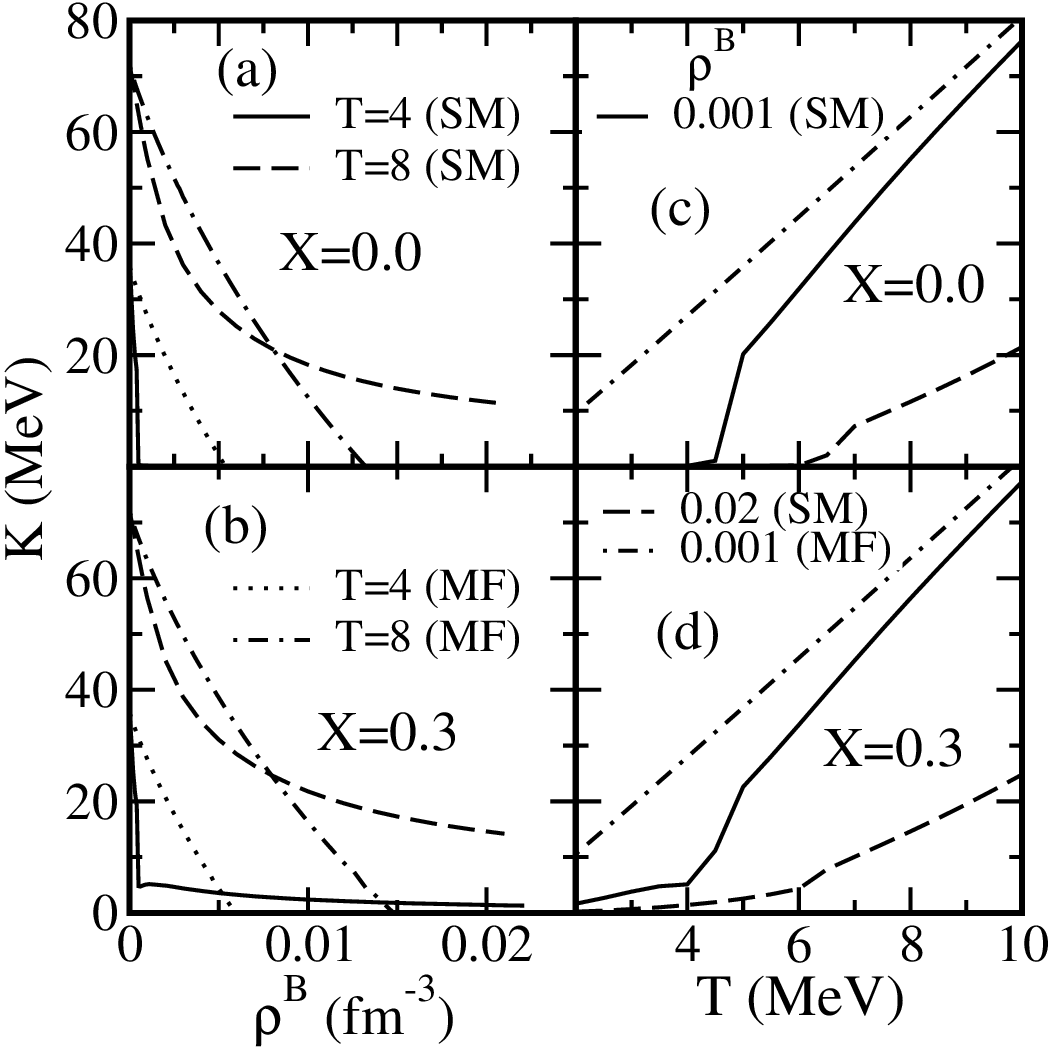}
\caption{ The incompressibility coefficient $K$ shown as a function
of the baryon density at $T=$4 and 8 MeV for symmetric 
and asymmetric nuclear matter (left panels) in the SM and MF models.
In the right panels $K$ is shown as a function of temperature
in the two models at the baryon densities indicated.}
\end{figure}

\begin{figure}
\includegraphics[width=1.2\columnwidth,angle=0,clip=true]{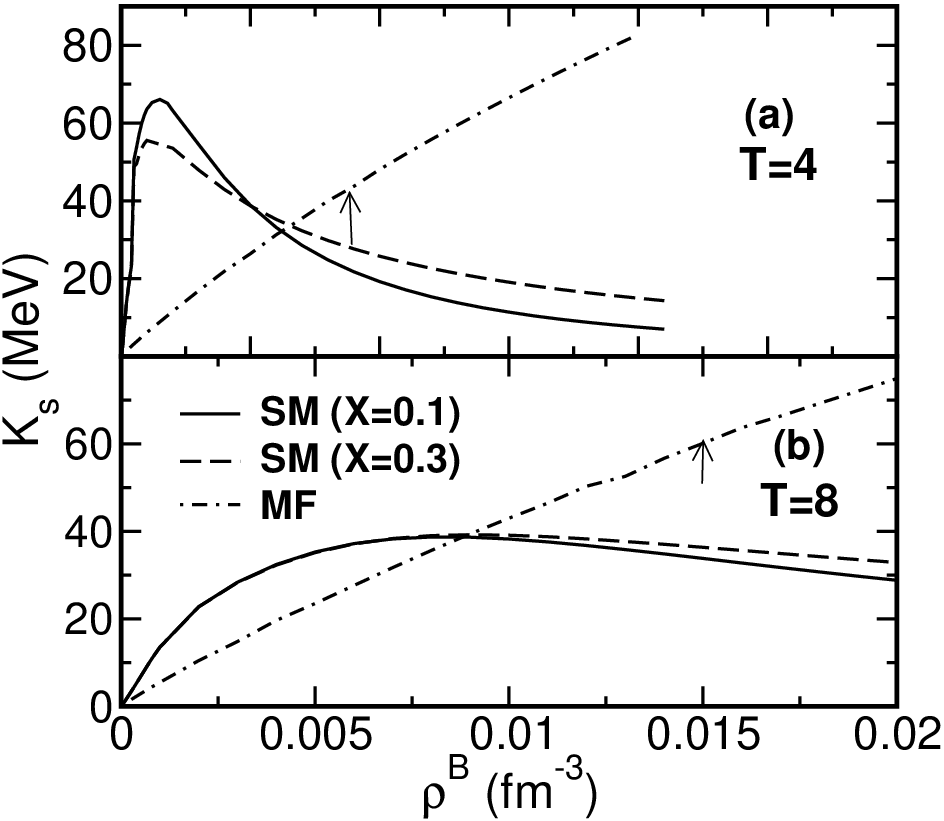}
\caption{ The symmetry incompressibility coefficient $K_s$
shown as a function of the baryon density at $T$= 4 MeV (upper panel)
and $T$= 8 MeV (lower panel) for asymmetric nuclear matter
with $X$= 0.1 and 0.3 in the SM and MF models. In the MF model,
$K_s$ is independent of asymmetry. The arrows denote the density
beyond which the system enters the unphysical region in the MF model.}
\end{figure}

\begin{figure}
\includegraphics[width=1.2\columnwidth,angle=0,clip=false]{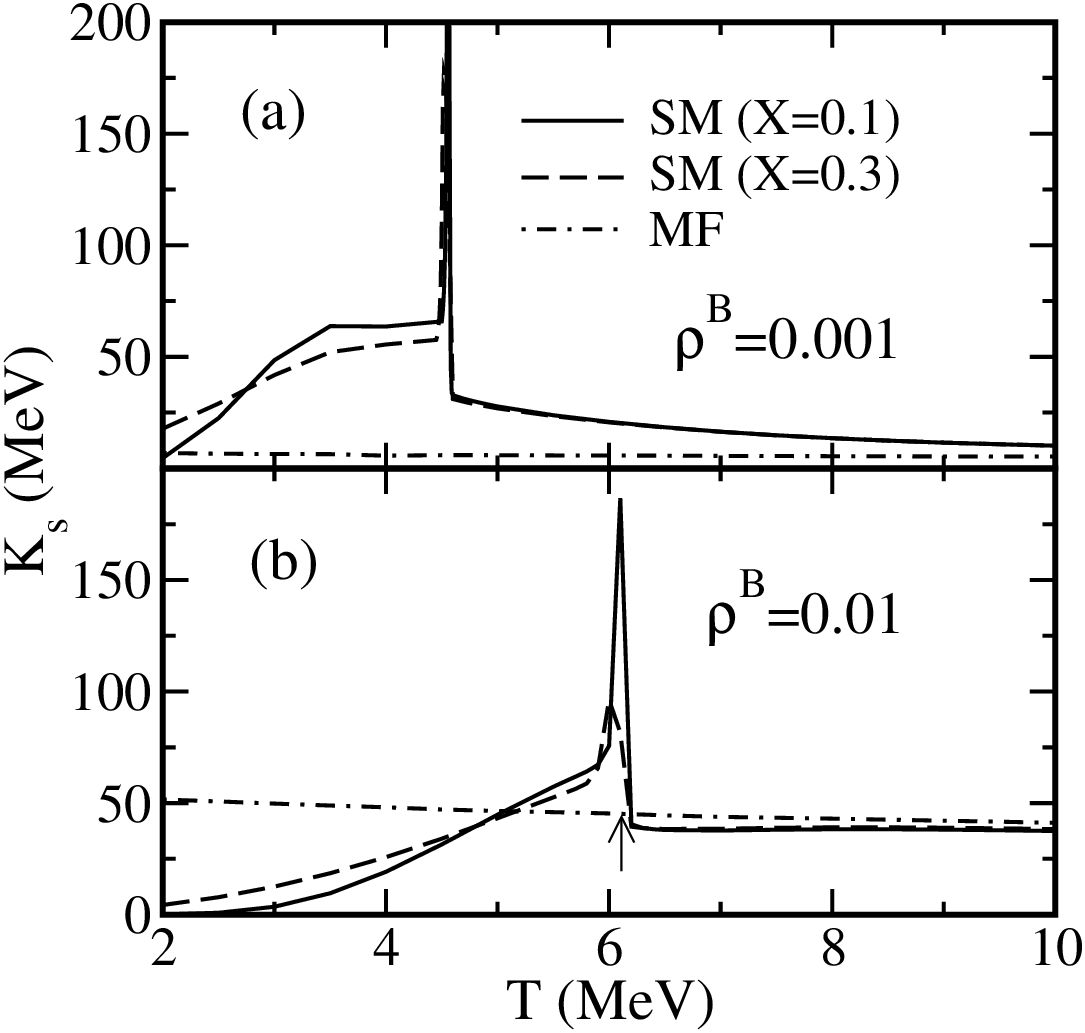}
\caption{The symmetry incompressibility $K_s$ for asymmetric nuclear
matter shown as a function of temperature at baryon densities
$\rho^B $=0.001 and 0.01 fm$^{-3}$ in the SM and MF
models. In the MF model, $K_s$ does not depend on asymmetry. 
The arrow in panel (b) denotes the temperature below which the
system enters the unphysical region in the MF model.}
\end{figure}

\begin{figure}
\includegraphics[width=1.2\columnwidth,angle=0,clip=true]{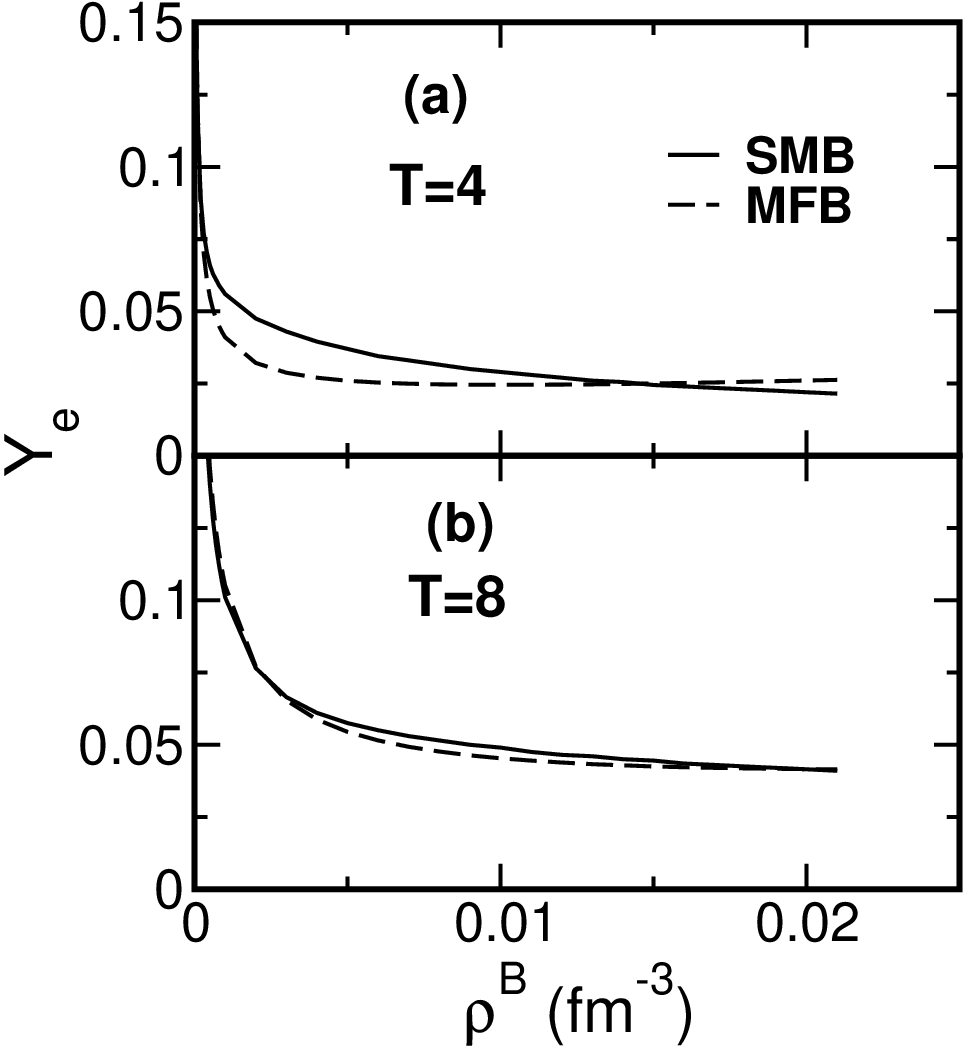}
\caption{The electron fraction $Y_e$ in $\beta$-equilibrated 
supernova matter as a function of the baryon density at $T$= 4 and
8 MeV in the $S$-matrix (SMB) and mean-field (MFB) models. }
\end{figure}

\begin{figure}
\includegraphics[width=1.2\columnwidth,angle=0,clip=true]{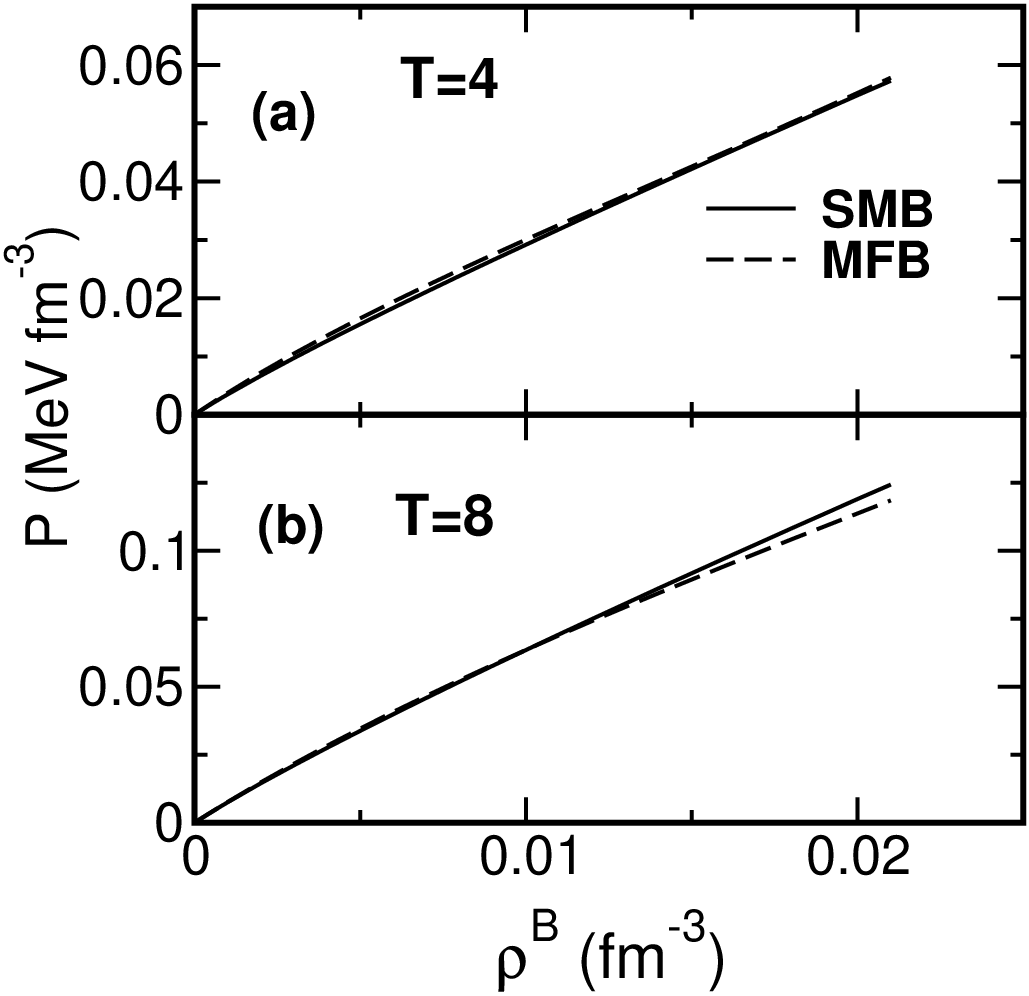}
\caption{The isotherms for $\beta$-equilibrated supernova
matter at $T$=4 and 8 MeV in the SMB and MFB models. }
\end{figure}

\begin{figure}
\includegraphics[width=1.2\columnwidth,angle=0,clip=true]{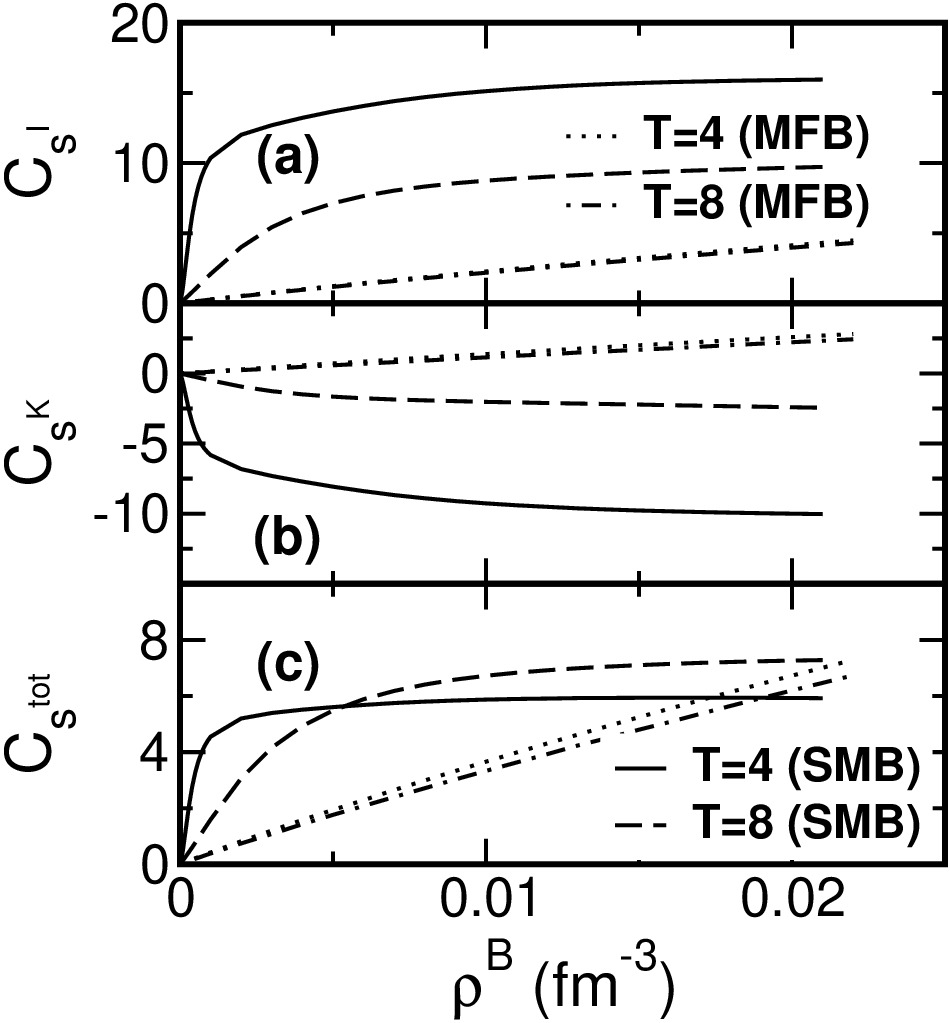}
\caption{The symmetry energy coefficients $C_s^I, C_s^K$,
and $C_s^{tot}$ (in MeV) of $\beta$-equilibrated supernova matter
as a function of the baryon density for $T$=4 and 8 MeV in
the SMB and MFB models. }
\end{figure}

\begin{figure}
\includegraphics[width=1.2\columnwidth,angle=0,clip=true]{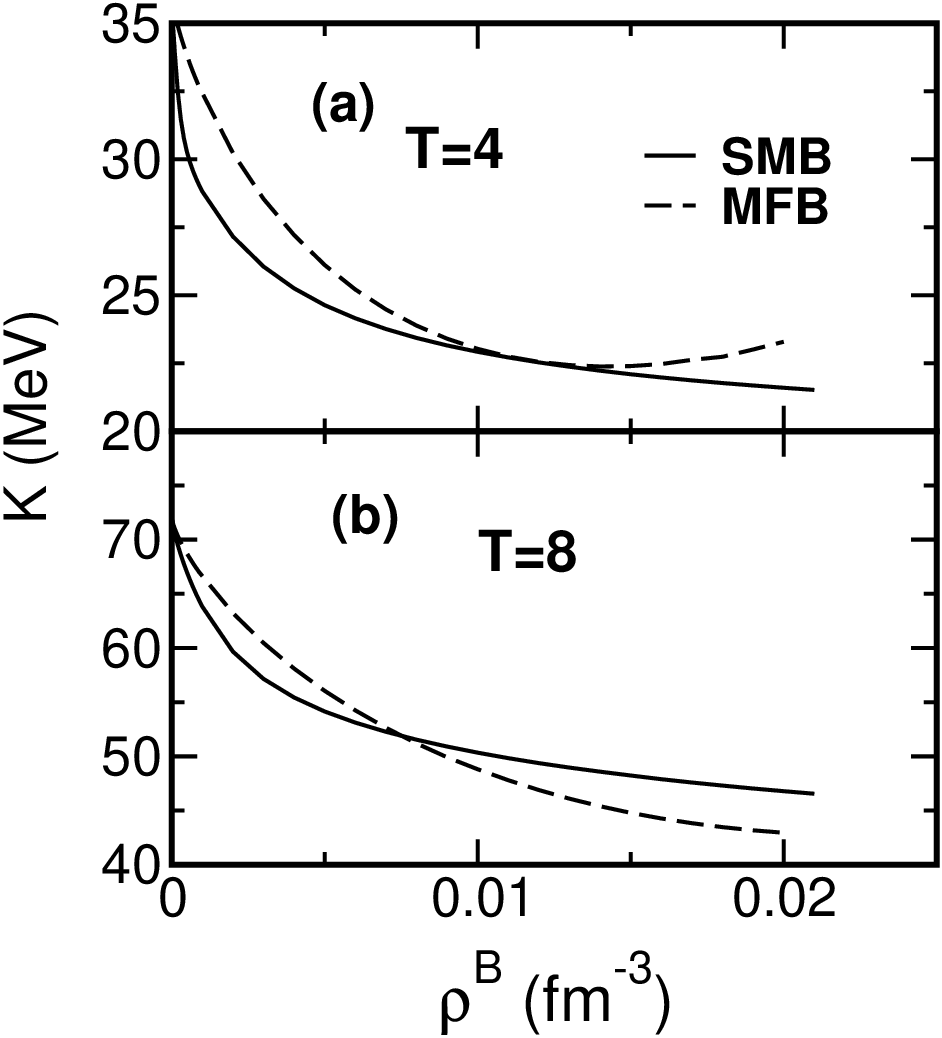}
\caption{The incompressibility of $\beta$-equilibrated supernova
matter as a function of the baryon density at $T$=4 and 8 MeV
in the SMB and MFB models. }
\end{figure}

\begin{figure}
\includegraphics[width=1.2\columnwidth,angle=0,clip=true]{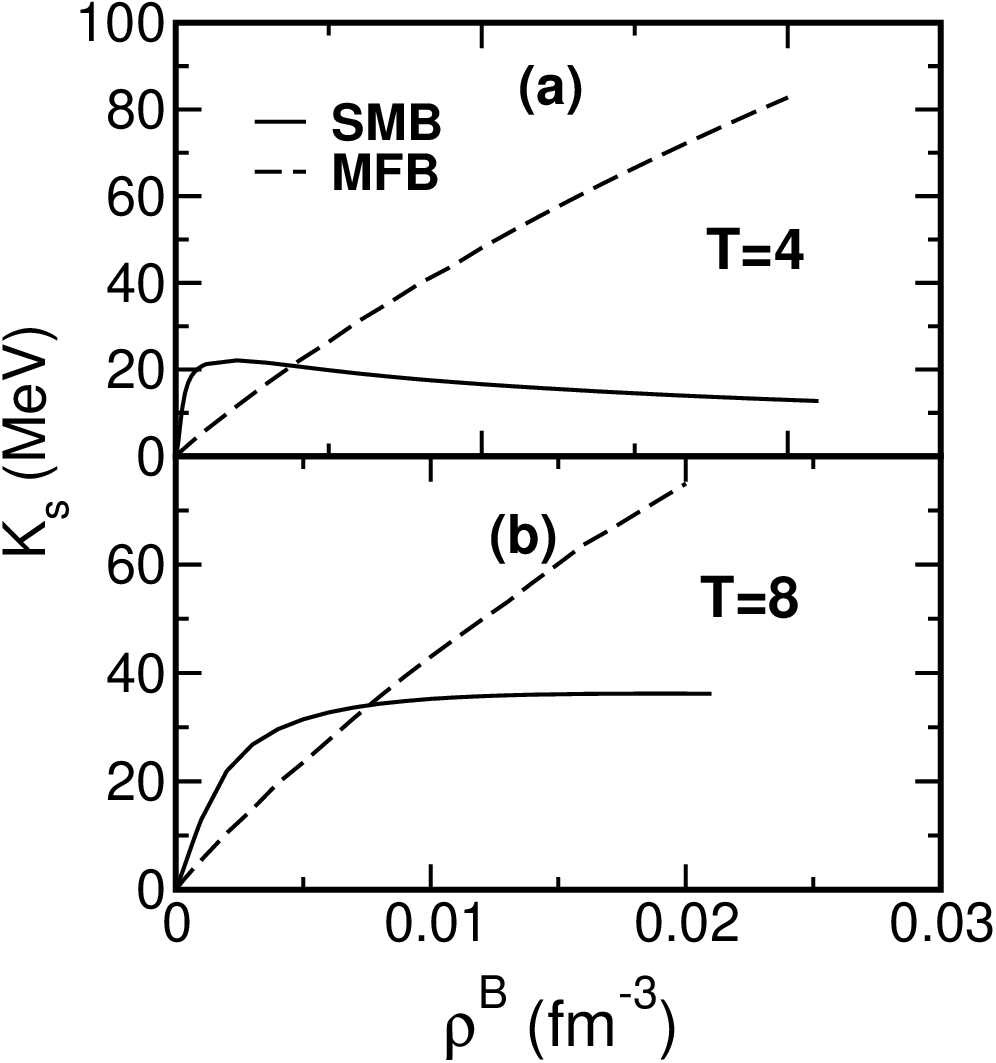}
\caption{The symmetry incompressibility  $K_s$ for supernova matter 
as a function of the baryon density at $T$=4 and 8 MeV in the SMB and
MFB models. }
\end{figure}

\end{document}